\documentclass{elsart}
\usepackage{graphicx}
\usepackage{amssymb}
\usepackage{subeqnarray}

\begin{document}
\begin{frontmatter}

\title{Phase transitions induced by microscopic disorder: a study based on the order parameter expansion}
\author{Niko Komin\corauthref{komin}},
\author{Ra\'ul Toral}

\address{IFISC (Instituto de F{\'\i}sica Interdisciplinar y Sistemas Complejos), Universitat de les Illes Balears-CSIC, Campus UIB, 07122 Palma de Mallorca, Spain}
\date{\today}
\corauth[komin]{Corresponding author.  {\sl Email address}: {\tt niko@ifisc.uib-csic.es}, Tel: +34-971-259520, Fax: +34-971-173426}

\begin{abstract}

Based on the order parameter expansion, we present an approximate method which allows us to reduce large systems of coupled differential equations with diverse parameters to three equations:  one for the global, mean field, variable and two which describe the fluctuations around this mean value. With this tool we analyze phase-transitions induced by microscopic disorder in three prototypical models of phase-transitions which have been studied previously in the presence of thermal noise. We study how macroscopic order is induced or destroyed by time independent local disorder and analyze the limits of the approximation by comparing the results with the numerical solutions of the self-consistency equation which arises from the property of self-averaging. 
Finally, we carry on a finite-size analysis of the numerical results and calculate the corresponding critical exponents.

\end{abstract}


\end{frontmatter}

\section{Introduction}
\label{cha:intro}

The effect of time-dependent noise in extended dynamical systems has been the subject of intensive study in the last years~\cite{GS:1999}. Besides the expected disordering role, it has been found that some kind of order at the macroscopic level can appear by increasing the intensity of the noise. Examples of this paradoxical result include stochastic resonance~\cite{srrmp,HM:2009}, or enhancement of the effect of an external forcing under the right amount of noise, coherence resonance \cite{PK:1997} (also named as stochastic coherence~\cite{ZGBUK:2003}) where a dynamical system displays optimal periodicity at the right noise value, noise sustained patterns, structures and fronts~\cite{SCSW:1997,CFT:2006}, phase transitions where a more ordered phase appears when increasing the noise intensity~\cite{BPT:1994,BPTK:1997}, etc.

In a very general framework, it has been argued that the resonance with an external forcing can also be achieved when the time-dependent noise is replaced by a more general source of disorder. This includes natural diversity or heterogeneity, competitive interactions, disorder in the network of connectivities, etc. and can appear in driven bistable and excitable systems~\cite{tessone2006,tessone2007a,ToralTessoneLopes2007}, in linear~\cite{toral2008a} and chaotic~\cite{chen2008} oscillators and in a variety of other systems~\cite{Gosak2009,Ullner2009,Zanette:2009,Tessone2009,Postnova09,Chen09,Wu09,Perc08,Acebron07}). A unifying treatment of the role of noise and diversity for non-forced excitable systems, has been developed in~\cite{TSTP:2007}. 
 
In this work we examine the effect that structural disorder or diversity, in the form of quenched noise, has on some prototypical models of phase transitions which have been thoroughly studied in the presence of noise. From the practical point of view, the models we will be considering bear some similarities with random-field, or impurities, models. As tool of investigation we will refine a previously developed order parameter expansion method of approximating large systems of coupled differential equations \cite{deMonteOvidio2002,deMonteOvidioMosekilde2003,deMonteOvidioChateMosekilde2004,deMonteOvidioChateMosekilde2005,iacyel2006} with diverse parameters. This allows the reduction of  the large set of differential equations to just  three: one for the global, mean field, value and two which describe the fluctuations around this mean value. Within this approximation we will analyze three different models and show its ease to deliver some understanding of the emergent properties of the global behavior. To find the limits of the order parameter expansion method we will compare the results with the solution of the self-consistency equation which arises from the property of self-averaging.

The chosen models are a set of globally coupled $\phi^4$-systems both in the presence of  additive and multiplicative quenched noise and the canonical model for noise-induced phase transition \cite{BPT:1994,BPTK:1997}. It will be seen that quenched noise can induce phase transitions (in, out and reentrant) of ordered phases.

The rest of the paper is organized as follows: In the next section we will describe the analytical methods, self-consistency and order parameter expansion; in section~\ref{cha:examples} we will apply those methods to the showcase models and compare with the results of numerical simulations; the last section closes the paper with the discussion of the presented results.

\section{Models and method}
\label{sec:modelAndMethod}

The type of models we will be considering in this work are defined via differential equations for the dynamics of a set of real variables $x_i$:
\begin{equation}
\label{eq:ODE_general}
   \dot x_i=f(x_i,\eta_i;X) \,, \hspace{2.0cm}i=1,\dots,N.
\end{equation}
The time derivative $\dot x_i(t)=dx_i(t)/dt$ depends on the constant parameter $\eta_i$, a kind of quenched noise. The set of values $\{\eta_1,\dots,\eta_N\}$ are independently drawn from a probability distribution $g(\eta)$ of mean $H$ and variance $\sigma^2$. Coupling between the different dynamical equations is provided by the presence in Eq.~(\ref{eq:ODE_general}) of the global variable or mean value $X(t)=\langle x_i(t)\rangle\equiv\frac{1}{N}\sum_{i=1}^N x_i(t)$. For a given realization of the $\eta_i$'s variables the $x_i$'s tend in the limit $t\to \infty$ to some asymptotic, stationary values which, in general, will depend on initial conditions. Some insight can be obtained if we write Eq.~(\ref{eq:ODE_general}) as a relaxational dynamics~\cite{msmtoral00} in a potential $V(x_i,\eta_i;X)=-\int^{x_i} d{x_i}' f({x_i}',\eta_i;X)$:
\begin{equation}
\label{eq:potential}
   \dot x_i=-\frac{\partial V(x_i,\eta_i;X)}{\partial x_i}.
\end{equation}
If the potential $V(x_i,\eta_i;X)$ is monostable for a particular value of $X$, then the variable $x_i(t)$ tends during the dynamical evolution towards the single minimum of $V(x_i,\eta_i;X)$. Note that the location of this minimum will change with time as $X$ evolves. If, on the contrary, $V(x_i,\eta_i;X)$ presents several minima, the dynamics will tend towards one of the local minimum of the potential.

In the following we will be interested in characterizing the stationary solution by the ensemble average value and fluctuations with respect to realizations of the quenched noise and initial conditions of the global variable $X$. We first review briefly the self-consistency method and then explain the approximate method based on the order parameter expansion.

\subsection{Self-consistency}
\label{cha:methodExact}

This method uses ideas borrowed from the Weiss molecular field theory~\cite{stanley}, which is known to be exact for systems with long-range interaction or, equivalently,  in which the interaction occurs through the global variable $X$, a mean-field scenario, as it is our case. Let us denote by $x_i^*$ the stable stationary solution of Eq.~(\ref{eq:ODE_general}). This is nothing but the absolute minimum of the potential $V(x_i,\eta_i;X)$. It will be a function of $\eta_i$ and the global variable $X$, i.e. $x_i^*=x^*(\eta_i,X)$. For a given realization of the quenched noise variables $\eta_i$'s, the value of the global variable must be obtained from the self-consistency relation $X=\frac{1}{N}\sum_{i=1}^Nx^*(\eta_i,X)$. It is clear that for $N$ large, the sum can be replaced by an integral over the distribution $g(\eta)$ of the independent $\eta_i$'s variables: 

\begin{equation}
\label{eq:selfCons}
 X=\int{d\eta g(\eta)x^*(\eta,X)}\,.
\end{equation}
It is then assumed that one can identify the value of $X$ obtained solving this equation, as the desired ensemble average, the self-averaging property~\cite{landaubinder}. In general, the possible solutions $X$ of the self-consistency equation (\ref{eq:selfCons}) have to be found numerically.  A possible scenario is that by changing some parameter (e.g. the root-mean square $\sigma$ or the mean $H$) of the distribution $g(\eta)$, the solutions bifurcate and the system then presents a phase transition between the possible solutions. We will present in the examples below the results of this procedure, but will not give any further details about the (in general, very involved) numerical method used to solve Eq. (\ref{eq:selfCons}).

\subsection{Order parameter expansion}
\label{cha:method}
For the development of this approximate method we assume, as in the previous subsection,  that the number of degrees of freedom $N$ is very large and then it is possible to substitute the mean value of the distribution $g(\eta)$ by the system average $H=\langle \eta_i \rangle =\frac{1}{N}\sum_{i=1}^N\eta_i$, the variance by  $\sigma^2=\langle (\eta_i-\langle \eta_i\rangle)^2\rangle=\frac{1}{N}\sum_{i=1}^N(\eta_i-\langle \eta_i\rangle)^2$, and similar expressions for other cases. 

Our goal is to find an approximate equation describing the dynamics of the mean value  variable $X$. To this end, we will expand the evolution equations in the deviations $\epsilon_i(t)=x_i(t)-X(t)$ of the dynamical variables from the mean value,  and the deviations $\delta_i=\eta_i-H$ of the parameters from their mean value. The Taylor expansion of Eqs.~(\ref{eq:ODE_general}) around the mean values up to second order gives:
\begin{eqnarray}
 \label{eq:ODE_taylored}
 \dot x_i &=& f(X,H;X)+\epsilon_i\,f_x(X,H;X)+\delta_i\,f_\eta (X,H;X)+\nonumber\\
&&	\frac{1}{2}\,\epsilon_i^2\,f_{xx}(X,H;X)+\epsilon_i\delta_i\,f_{x\eta}(X,H;X)+\frac{1}{2}\,\delta_i^2\,f_{\eta\eta}(X,H;X)+\dots
\end{eqnarray}
With the usual notation $f_x(X,H;X)=\left.\frac{\partial f(x,\eta;X)}{\partial x}\right|_{x=X,\eta=H}$, etc. We now take averages and use that $\langle\epsilon_i\rangle=\frac{1}{N}\sum_{i=1}^N\epsilon_i=0$ and $\langle\delta_i\rangle=\frac{1}{N}\sum_{i=1}^N\delta_i=0$. Furthermore we have $\langle\delta_i^2\rangle=\sigma^2$ as the parameter distribution's variance. So when we average over Eq.~(\ref{eq:ODE_taylored}) we are left with:
\begin{eqnarray}
 \label{eq:ODE_meanX}
\dot{X} &=&f(X,H;X)+\frac{1}{2}f_{xx}(X,H;X)\langle\epsilon_i^2\rangle+f_{x \eta}(X,H;X)\langle\epsilon_i\delta_i\rangle+\nonumber\\
&&\frac{\sigma^2}{2}f_{\eta\eta}(X,H;X)+{ O}(\langle\epsilon_i^3\rangle,\langle\epsilon_i^2\delta_i\rangle,...) \, .
\end{eqnarray}
The evolution of $X$ is then coupled to that of the second moment of the {\it snapshot probability density} $\Omega=\langle\epsilon_i^2\rangle=\frac{1}{N}\sum_{i=1}^N\epsilon_i^2$ and the so-called {\it shape parameter}~\cite{deMonteOvidioChateMosekilde2005}  $W=\langle\epsilon_i\delta_i\rangle=\frac{1}{N}\sum_{i=1}^N\epsilon_i\delta_i$.  We will now obtain evolution equations for these two variables. We follow closely the method of \cite{deMonteOvidioMosekilde2003} but keeping all terms up to second order in $\epsilon_i$ and $\delta_i$.  We start by subtracting~(\ref{eq:ODE_meanX}) from~(\ref{eq:ODE_taylored}) to obtain $\dot \epsilon_i=\dot x_i - \dot{X}$, which can then be replaced in $\dot \Omega=\langle 2 \epsilon_i\dot\epsilon_i\rangle$, $\dot W=\langle \delta_i\dot \epsilon_i\rangle$. After some algebra, and neglecting terms of order ${ O}(\langle\epsilon_i^3\rangle,\langle\epsilon_i^2\delta_i\rangle,\dots)$ or higher, we get:
\begin{subeqnarray}
\label{eq:OPE_system}
\slabel{eq:ODE_X}
\dot X &=&f(X,H;X)+\frac{\Omega}{2} f_{xx}(X.H;X)+f_{x \eta}(X,H;X)\,W+\nonumber\\
& & \frac{\sigma^2}{2}f_{\eta\eta}(X,H;X),\\
\slabel{eq:ODE_omega}
 \dot \Omega&=& 2\,\Omega\,f_x(X,H;X)+2\,W\,f_\eta(X,H;X),\\
\slabel{eq:ODE_W}
 \dot W&=& W\,f_x(X,H;X)+\sigma^2f_{\eta}(X,H;X).
\end{subeqnarray}
In summary, within this approximation we have obtained a closed set of three differential equations~(\ref{eq:ODE_X}-\ref{eq:ODE_W}). They have the feature of being coupled only in one direction, i.e. $W(t)$ is independent of the others and $\Omega(t)$ depends only on $W(t)$. Steady states with $\dot W=\dot \Omega=\dot X=0$ are $\displaystyle W=-\sigma^2\frac{f_{\eta}}{f_{x}}$ and $\displaystyle \Omega=\sigma^2\frac{f_{\eta}^2}{f_x^2}$. The equilibrium of variable $X$ is given by the solution of:
\begin{equation}
 \label{eq:equilibrium_condition}
0=f+\frac{\sigma^2}{2}\left[f_{\eta\eta}+f_{xx}\frac{f_\eta^2}{f_x^2}-2\frac{f_{x\eta}\,f_{\eta}}{f_x}\right]\,,
\end{equation}
where we have simplified notation $f=f(X,H;X)$, etc. As before, an analysis of the bifurcations of this equation will allow us to find the possible phase transitions of the model.

Our results, Eqs.(\ref{eq:OPE_system}), differ slightly from those in the cited sources. In \cite{deMonteOvidio2002,deMonteOvidioChateMosekilde2004,deMonteOvidioChateMosekilde2005} the authors require the parameter to be additive, thus setting $f_{\eta}=1$. In other works \cite{deMonteOvidioMosekilde2003,iacyel2006} any parameter dependence is allowed, but a coherent regime is required, such that terms of order ${O}(\langle\epsilon_i^2\rangle)$ and higher are neglected.

\section{Examples}
\label{cha:examples}
After presenting the general development of the order parameter expansion method, we will now apply it to a few models of relevance in the field of phase transitions. Our purpose is to compare the results of our approximation with those coming from the self-consistency equation analysis as well as with numerical simulations of the different models. Solving the self-consistency equation requires in practice a complicated numerical calculation, while our treatment is simple and predicts in some cases the existence of phase transitions with reasonable accuracy.

\subsection{Globally coupled Landau-Ginzburg model with additive quenched noise}
\label{cha:landauAdditive}
The Landau-Ginzburg or $\phi^4$ scalar-field has been thoroughly studied from the analytical and numerical points of view as a paradigmatic model undergoing a second-order phase transition~\cite{amit}. Here we are interested in this model in the case that the stochastic thermal fluctuations have been replaced by additive quenched noise, as an example of a random-field scalar model~\cite{young}. The dynamical equations for the set of $x_i, i=1,\dots,N$, real variables are:
\begin{equation}
 \label{eq:ODE_landauAdditive}
\dot x_i=a\,x_i-x_i^3+C\left(X-x_i\right)+\eta_i \,.
\end{equation}
The study of the model using the self-consistency relation Eq.~(\ref{eq:selfCons}) can be found in~\cite{ToralTessoneLopes2007}. Here we want to use the order parameter expansion to derive the main properties of this model, in particular the existence of a phase transition as a function of the intensity $\sigma$ of the fluctuations of the random fields $\eta_i$.

Following the steps from section \ref{sec:modelAndMethod}, we obtain the set of equations for the order parameter $X$ and the fluctuations $W,\Omega$:
\begin{subeqnarray}
\label{eq:OPE_phiadd_system}
 \slabel{eq:OPE_landauAdd_meanX}
\dot{X}&=&\left(a-3\,\Omega\right)\,X-X^3+H\\
\slabel{eq:OPE_landauAdd_omega}
\dot \Omega&=&2\,\Omega\left(a-C-3\,X^2\right)+2\,W\\
\slabel{eq:OPE_landauAdd_w}
\dot W&=&W\,\left(a-C-3\,X^2\right)+\sigma^2 \,.
\end{subeqnarray}
The steady state for the order parameter, Eq.~(\ref{eq:equilibrium_condition}), leads to:
\begin{equation}
 \label{eq:OPE_landauAdd_equiCond}
0=\left( a-3\,{\frac {{\sigma}^{2}}{ \left(3\,{X}^{2}+C -a
 \right) ^{2}}} \right) X-{X}^{3} +H\, .
\end{equation}
We now consider the case of zero average field $H=\langle\eta_i\rangle=0$. In that case, Eq.~(\ref{eq:OPE_landauAdd_equiCond}) can have up to five real solutions. The trivial solution  $X=0$, always exists and it is stable (if $C>a$) whenever $\sigma>\sigma_c$, with

\begin{equation}
\label{eq:phiAdd_sigmaCrit}
 \sigma_{c}=\cases{ 0 &  if $a<0$,\cr
 \sqrt{\frac{a}{3}}(C-a) & if $a>0$.
 }
\end{equation}

It turns out that for $C>7a>0$, the set of Eqs. (\ref{eq:OPE_landauAdd_meanX}-{\ref{eq:OPE_landauAdd_w}) contains two additional stable fixed point real solutions $\pm X_0$ for $\sigma\le \sigma_c$. At $\sigma=\sigma_c$ it is $X_0=0$ and hence $\sigma_c$ identifies a second order, continuous, phase transition (see right panel of Fig. \ref{fig:phiAdd_bifur}). If $7a>C>a>0$ the range of existence and stability of these two additional solutions extends up to $\sigma\le \sigma_0$, where $\sigma_0\ge\sigma_c$ is given by:
\begin{equation}
 \label{eq:phiAdd_sigmaCrit2}
 \sigma_0=\sqrt{\frac{4}{243}\left(2\,a+C\right)^{3}},
\end{equation}
Hence, in the range $\sigma\in[\sigma_c,\sigma_0]$ there is bistability between the $X=0$ and the $\pm X_0$ solutions. Moreover, two additional symmetric unstable solutions $\pm X_1$ appear in this range. Therefore, the point $\sigma_0$ signals the appearance of a first order, discontinuous, phase transition (see Fig.~\ref{fig:phiAdd_bifur}, \textit{left}). In that range, the three stable solutions coexist with the two unstable solutions.

From a microscopic point of view, the phase transition from the $|X|>0$ to the $X=0$ states can be explained as follows: for $\sigma=0$, it is $\eta_i=0, \forall i$; all variables end up in the same stationary value $x_i=\sqrt{a}$ or $x_i=-\sqrt{a}$ and the average value satisfies $|X|=\sqrt{a}>0$ . As the noise intensity increases, $\sigma>0$, the average value $|X|$ tends to zero and the chances that individual values $\eta_i$ are smaller than $-CX$ grow. This changes the minimum's sign in the (individual) potential. As a consequence the distribution of $\{x_i\}$ becomes bimodal and the mean value approaches zero.

\begin{figure}[h]
 \centering
 \includegraphics[width=5cm]{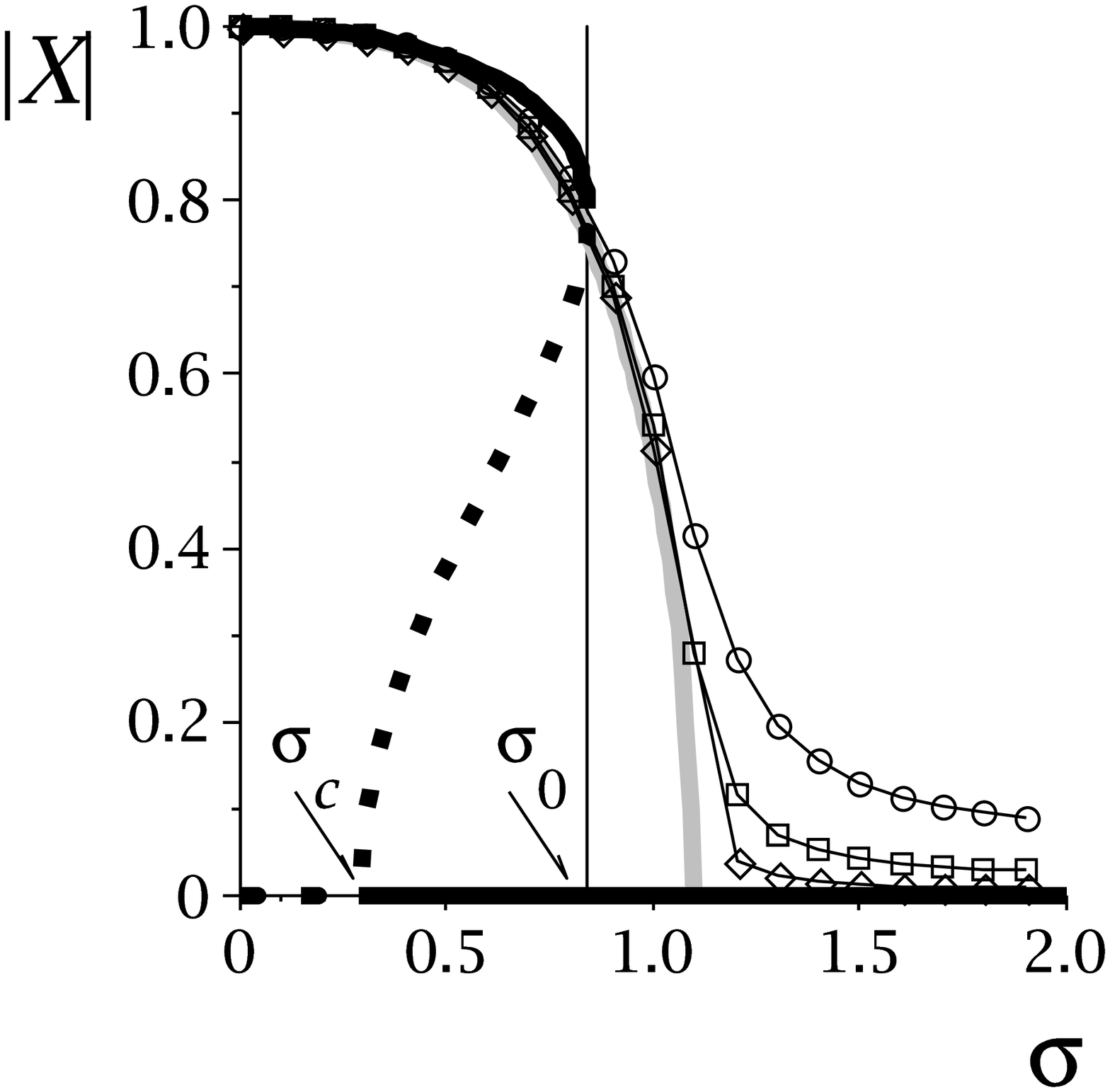}
 \includegraphics[width=5cm]{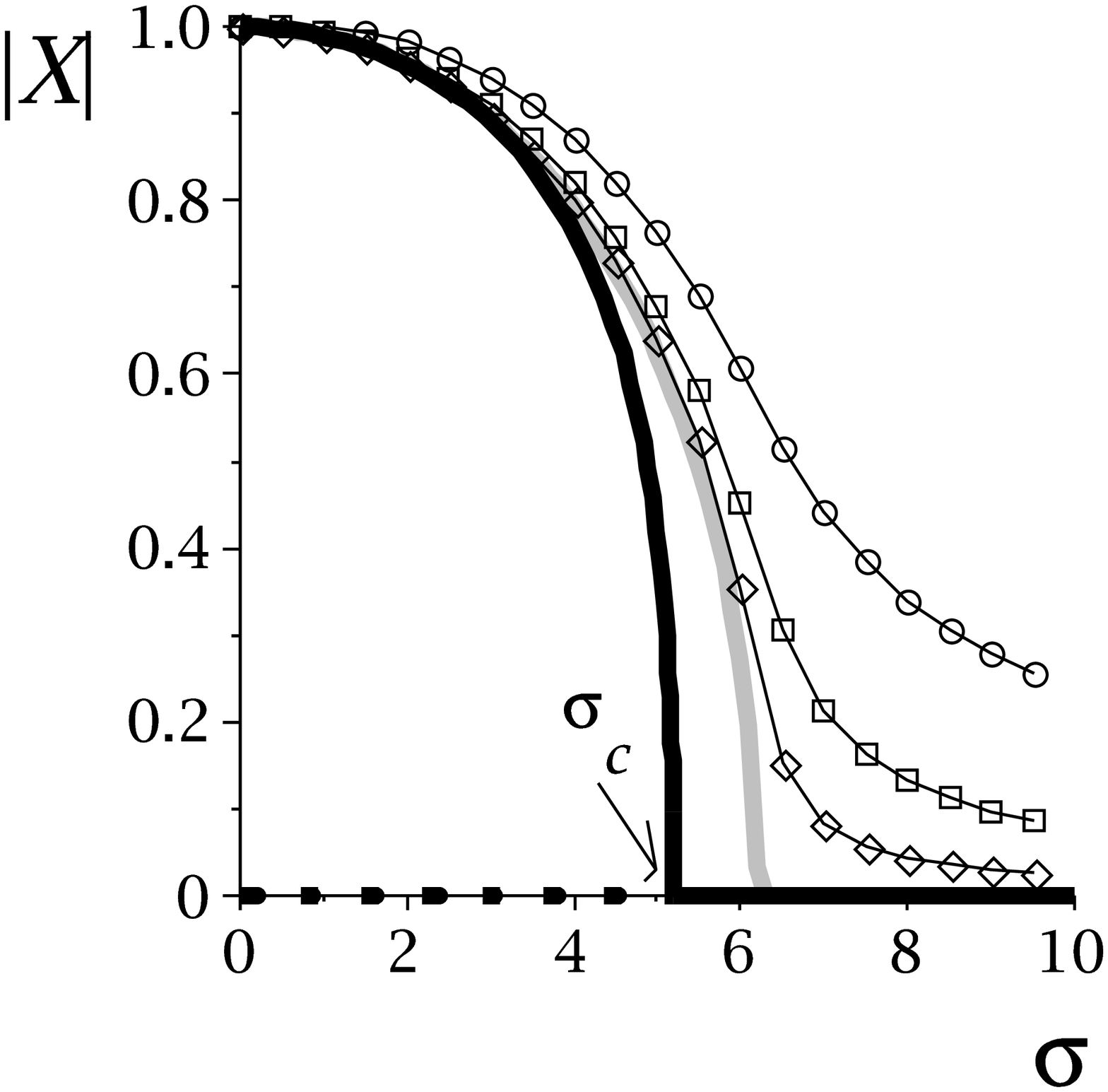}
 \caption{Bifurcation diagram of the Landau-Ginzburg model with additive quenched noise. Order parameter expansion (thick black lines) predict a second order transition for $C>a$ (\textit{right}: $C=10,a=1$) while bistability (first order phase transition) appears for $C<7a$ (\textit{left}: $C=1.5, a=1$, the unstable solution is plotted as a dotted line).  The self-consistency solution (grey line) does not show bistability in any case. Symbols show the results of numerical simulations of the evolution equations averaged over $10^3$ realizations of the  quenched noise variables $\eta_i$ and initial conditions. $N=10^3,10^4,10^5$ (circles, squares, diamonds, respectively).}
 \label{fig:phiAdd_bifur}
\end{figure}

The existence of a phase transition from order to disorder predicted by the order parameter expansion simple approximation scheme is confirmed by the numerical solution of the self-consistency equation (\ref{eq:selfCons}) \cite{ToralTessoneLopes2007}. However, the transition appears to be always second-order, so indicating the validity of the prediction of the approximate order parameter expansion in the limit of large coupling. In fact, the critical value $\sigma_c$ predicted by the order-disorder transition, Eq. (\ref{eq:phiAdd_sigmaCrit}), deviates systematically from that coming from the numerical integration of the self-consistency equation (\ref{eq:selfCons}) for large values of the coupling constant $C$, as shown in figure \ref{fig:phiAdd_sigmaDeC}, although the relative error between the two values decreases as $C$ increases.

\begin{figure}[h]
 \centering
 \includegraphics[width=5cm]{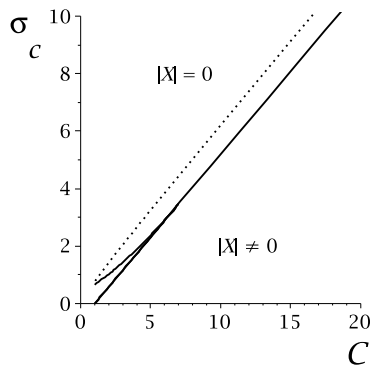}
 \caption{Critical intensity of the additive quenched noise for the Landau-Ginzburg model versus coupling strength for $a=1$. Prediction of order parameter expansion~(\ref{eq:phiAdd_sigmaCrit}) as continuous line, exact solution (\ref{eq:selfCons}) as dotted line. The order parameter expansion predicts a bistability region for $C<7a$.}
 \label{fig:phiAdd_sigmaDeC}
\end{figure}

We have also compared these predictions versus the results coming from intensive numerical simulations. In the simulations we have integrated the full set of equations (\ref{eq:ODE_landauAdditive}) up to the steady state and, then, we have computed the order parameter $m=\langle\langle |X| \rangle\rangle$ and  its fluctuations $\chi=\frac{N}{\sigma^2}\, \left[\left\langle\langle X^2\right\rangle\rangle-\left\langle\langle\vert X\vert\right\rangle\rangle^2\right]$. Here $X=\frac{1}{ N}\sum_{i=1}^Nx_i$ and $\langle\langle\cdots\rangle\rangle$ denotes an ensemble average with respect to realizations of the random variables $\eta_i$ and initial conditions. The simulation results for the order parameter are indicated by symbols in figure \ref{fig:phiAdd_bifur}. As usual, the transition from order to disorder is smeared out due to finite-size-effects but the numerical simulations do approach the results of the self-consistency equation as the number $N$ of variables increases. We have analyzed our data using standard finite-size-scaling relations~\cite{cardy,deutsch:92} and found that the dependence of the order parameter on $\sigma$ can be well fitted by $m(\sigma,N)=N^{-b/2}f_m(\epsilon N^{b})$ with $\epsilon=1-\sigma/\sigma_c$, $b\approx 0.33$ and being $f_m$ a scaling function, see evidence in the left panels of figure~(\ref{fig:phiAdd_simu}) for two different values of the coupling constant. Note that this scaling relation implies that in the thermodynamic limit, the order parameter vanishes as $m(\sigma)\sim(\sigma_c-\sigma)^{1/2}$, the typical mean-field result. Similarly, the fluctuations can be fitted by the form $\chi(\sigma,N)=N^{c}f_{\chi}\left(\epsilon N^{b}\right)$, with  $c\approx 0.67$ and $f_{\chi}$ the appropriate scaling function, as demonstrated in the right panels of  figure (\ref{fig:phiAdd_simu}) again for two different values of the coupling constant. This implies that in the thermodynamic limit, the fluctuations diverge as $\chi(\sigma)\sim|\sigma_c-\sigma|^{-\gamma}$ with $\gamma=c/b\approx 2$.

\begin{figure}[h]
 \centering
 \includegraphics[bb=50 50 276 276,width=6cm]{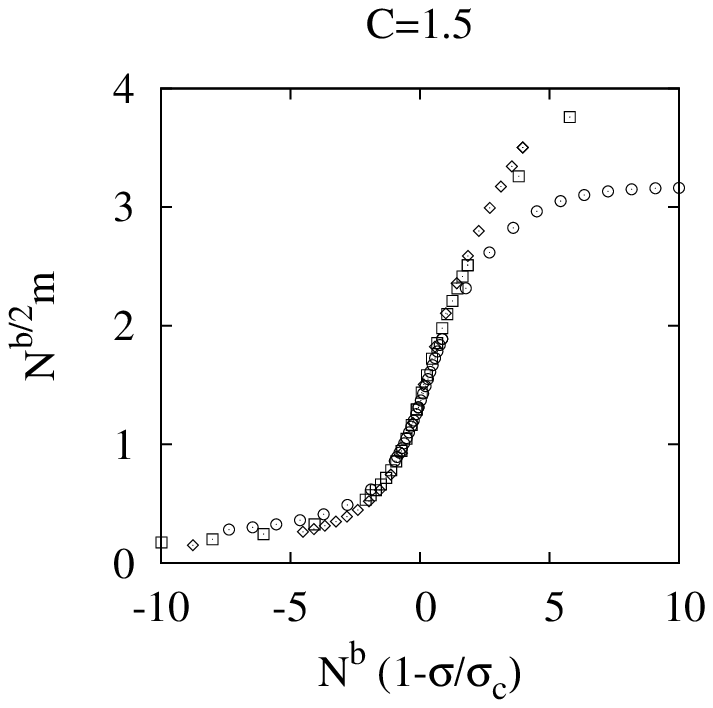}
 \includegraphics[bb=50 50 276 276,width=6cm]{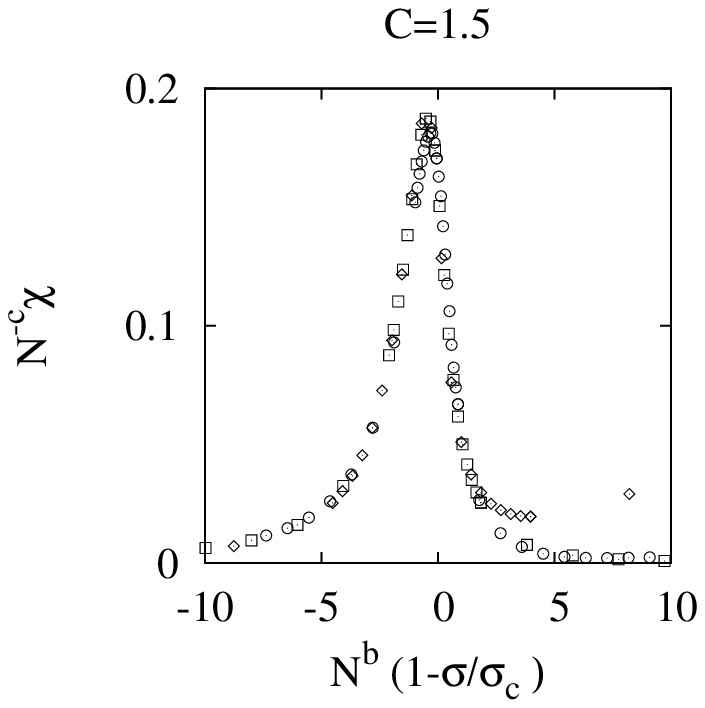}
 \includegraphics[bb=50 50 276 276,width=6cm]{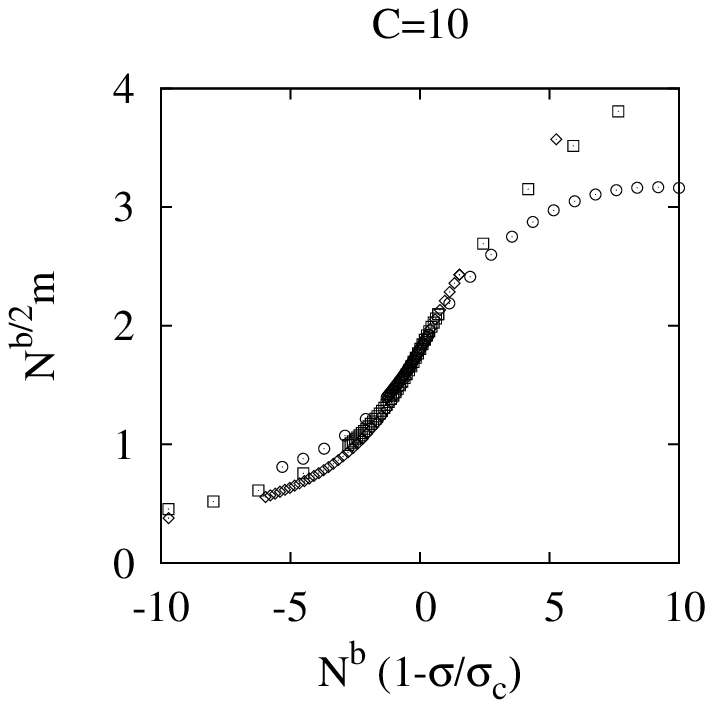}
 \includegraphics[bb=50 50 276 276,width=6cm]{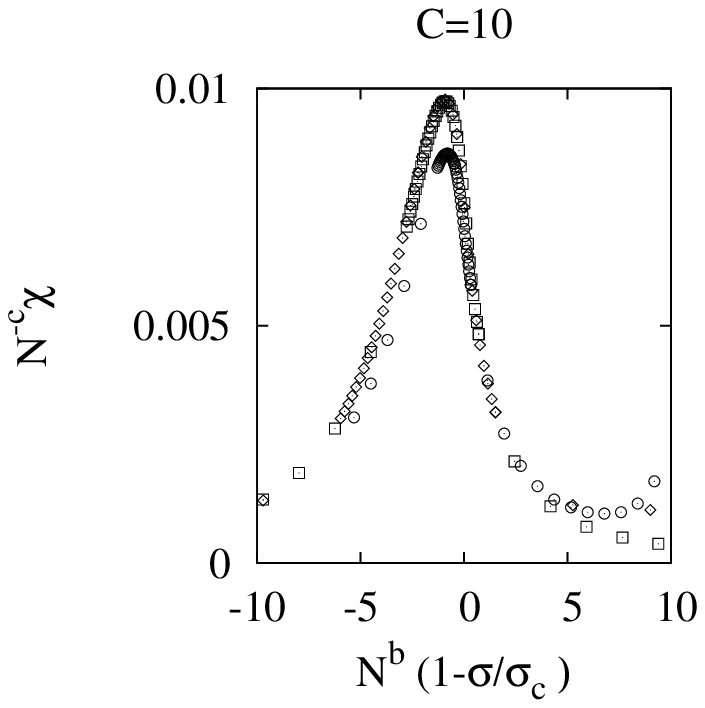}
 \caption{Finite-size scaling analysis of the Landau-Ginzburg model with additive quenched noise. Rescaled simulation data for low coupling (\textit{top graphs}, $\sigma_c=1.094$) and high coupling (\textit{bottom graphs}, $\sigma_c=6.203$). Ensemble average $m$  (\textit{left}) and fluctuations $\chi$(\textit{right}) as defined in the main text. Exponents: $b=0.33, c=0.67$. Ensemble sizes: $N=10^3, 10^4, 10^5$ (circles, squares, diamonds). In all cases: $a=1$. Here, $\sigma_c$ has been determined to a high degree of accuracy by using the numerical solution of Eq.~(\ref{eq:selfCons}).}
 \label{fig:phiAdd_simu}
\end{figure}

\subsection{Globally coupled Landau-Ginzburg model with multiplicative quenched noise}
\label{cha:landauMultiplicative}
We now consider the case in which the quenched noise couples multiplicatively to the variable $x_i$:
\begin{equation}
 \label{eq:ODE_landauMulti}
\dot x_i=\left(a+\eta_i\right)\,x_i-x_i^3+C\left(X-x_i\right)\,.
\end{equation}
This model has been extensively studied in the case that the $\eta_i$'s are independent white noises and it has been found that an increase in the noise intensity leads to a transition from disorder to order~\cite{GS:1999,VPAH94,GPSV96,BP:01}. We want to compare the predictions of the self-consistency equation with the order parameter expansion and numerical simulations to check if a similar result holds in the case of quenched noise. Without coupling ($C=0$) Eq.~(\ref{eq:ODE_landauMulti}) is a prototype of supercritical pitchfork bifurcations (see e.g. in~\cite{StrogatzNonlinearDyn}) with two possible sets of solutions: $x_i=0$ is the stable solution whenever $a+\eta_i\le0$ or   $x_i=\pm\sqrt{a+\eta_i}$ are stable solutions and $x_i=0$ is unstable for $a+\eta_i>0$.

To study the consequences of coupling, $C>0$, we use the above developed order parameter expansion approximation. After setting $H=\langle\eta_i\rangle=0$, the equations are:
\begin{subeqnarray}
\label{eq:OPE_phimult_system}
\slabel{eq:OPE_phi4Mult_X}
 \dot {X}&=& a X-X^3-3 X\Omega + W, \\
\slabel{eq:OPE_phi4Mult_O}
 \dot \Omega &=& \left(2 a - 6X^2-2C \right)\Omega+2X W,\\
\slabel{eq:OPE_phi4Mult_W}
 \dot W&=&\left(a -3X^2-C\right)W+X\sigma^2 \,.
\end{subeqnarray}
The equilibrium condition~(\ref{eq:equilibrium_condition}) leads to:
\begin{equation}
\label{eq:phiMult_equilibrium}
0=a\,X-{X}^{3}-\,{\frac {3{X}^{3}{\sigma}^{2}}{ \left( C-a+3\,{
X}^{2}\right) ^{2}}}+{\frac {X{\sigma}^{2}}{C-a+3\,{X}^{2}}} \, .
\end{equation}

Similarly to the uncoupled case this equation has two different regimes of solutions: 
On one hand,  if $a\ge0$  the stable solutions of Eq.~(\ref{eq:OPE_phimult_system}) are $X=\pm\sqrt a$ for $\sigma=0$. As $\sigma$ increases, $|X|$ monotonically increases as well (see fig.~\ref{fig:phiMult_bifur}, \textit{left}). On the other hand, if $a<0$ then $X=0$ is a stable solution for small $\sigma$. At some value $\sigma_c$ it becomes unstable and a fork of solutions grows out of zero (see fig.~\ref{fig:phiMult_bifur}, \textit{right}). $\sigma_c$ is determined by Eq.(\ref{eq:phiMult_equilibrium}) and is related to $a$ and $C$ by:
\begin{equation}
\sigma_c^2=a\left(a-C\right).
\end{equation}
$\sigma_c$ identifies a second-order phase transition from disorder to order (i.e. from $X=0$ to $X\neq0$). In this case of $a<0$, the value $\sigma_c$ grows monotonously with coupling strength $C$, a rather counterintuitive observation, since it means that the coupling hinders the ordering and more structural disorder is needed to induce macroscopic order (fig.~\ref{fig:phiMult_sigma_de_C}).

\begin{figure}[h]
 \centering
 \includegraphics[width=5cm]{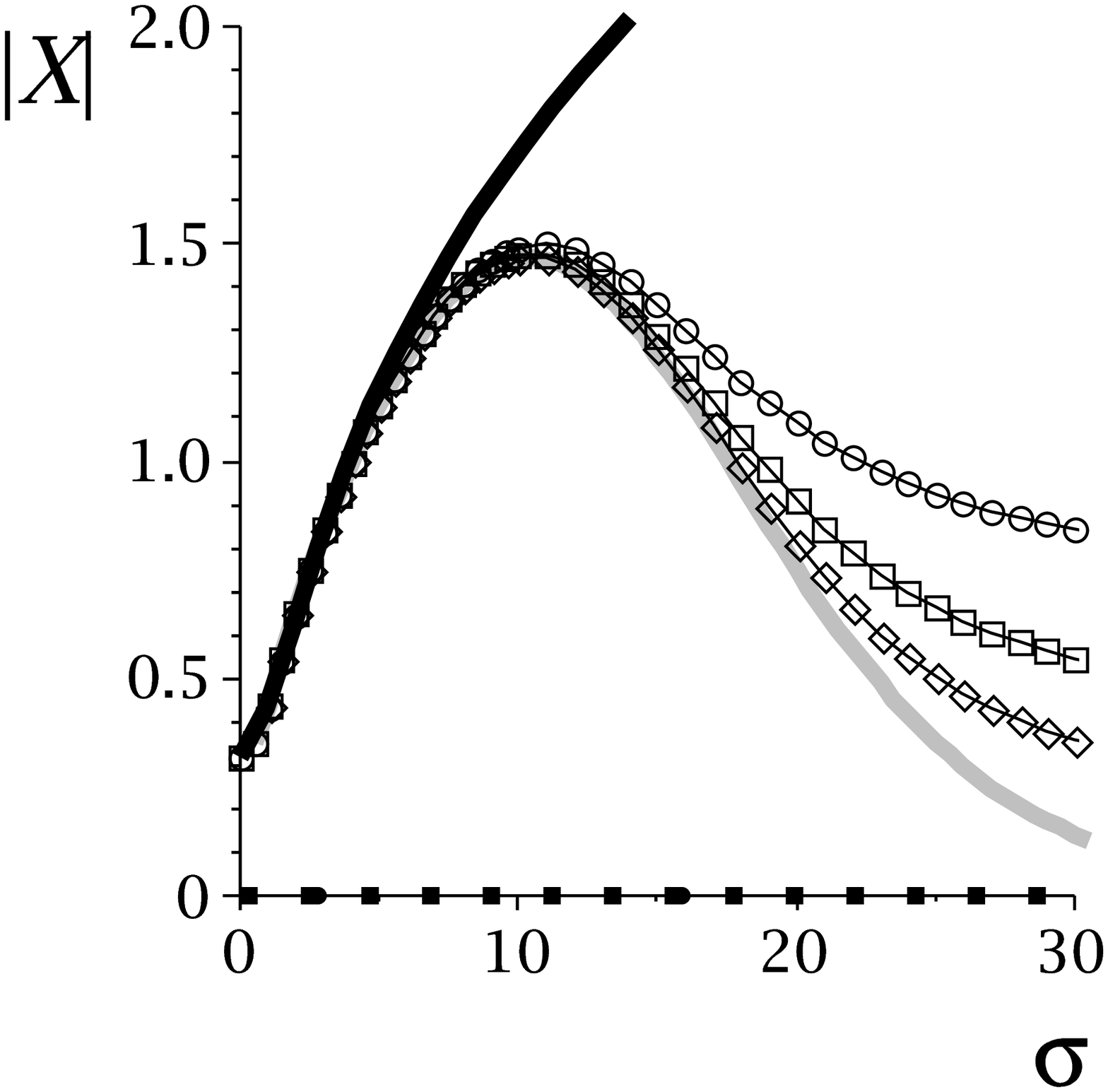}
 \includegraphics[width=5cm]{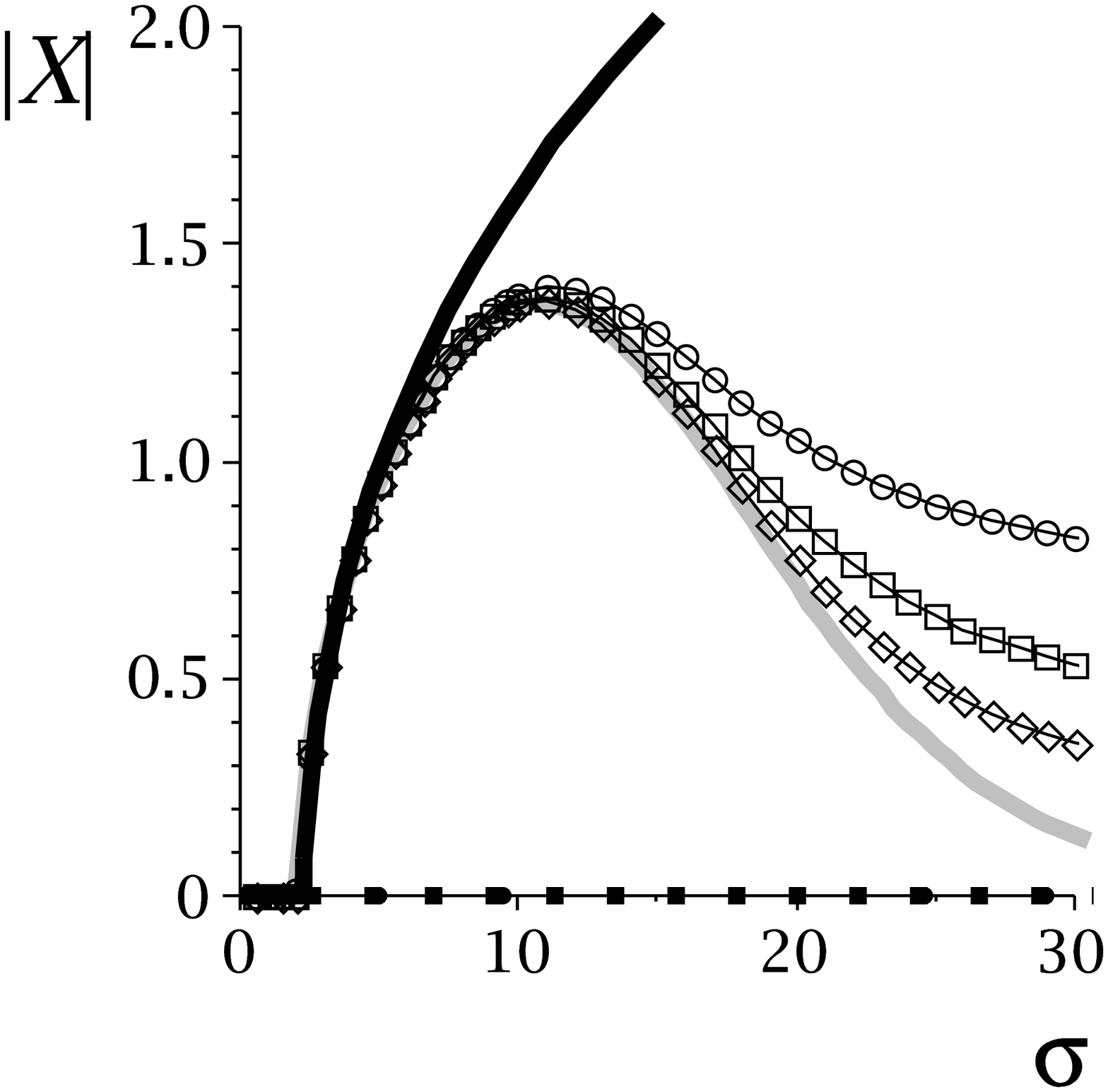}
 \caption{Bifurcation diagram of the Landau-Ginzburg model with multiplicative quenched noise. Positive values of $a$ show order without noise (\textit{left:} $a=0.1$), whereas negative value show order only with a finite value of the noise intensity (\textit{right:} $a=-0.5$). In both panels it is $C=10$. The order parameter expansion approximation scheme gives a monotonous solution while the exact solution of Eq.~(\ref{eq:selfCons}) reaches a maximum and decreases for large $\sigma$ (grey line). Symbols are the result of direct numerical simulations of Eqs.~(\ref{eq:ODE_landauMulti}) averaged over $10^3$ realizations of the quenched noise variables $\eta_i$ for $N=10^3,10^4,10^5$ (circles, squares, diamonds). 
 \label{fig:phiMult_bifur}}
\end{figure}
\begin{figure}[h]
 \centering
 \includegraphics[width=5cm]{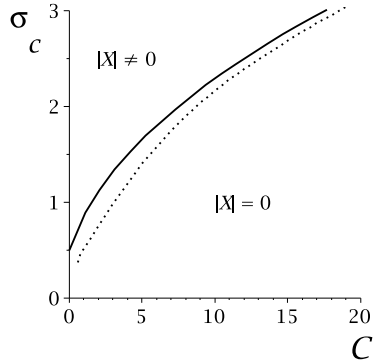}
 \caption{Critical noise for bifurcation versus coupling strength for $a=-0.5$ for the Landau-Ginzburg model with multiplicative quenched noise. The prediction by the order parameter expansion is shown as continuous line, the exact solution (\ref{eq:selfCons}) is shown as dotted line.}
 \label{fig:phiMult_sigma_de_C}	
\end{figure}

The numerical solution of the self-consistency equation (\ref{eq:selfCons}) is qualitatively similar to the results of the order parameter expansion approximation, however $|X|$ doesn't increase monotonically with increasing $\sigma$. It rather reaches a maximum and decreases after that approaching zero asymptotically. Note that his is not a (reentrant) phase transition since $|X|=0$ is only reached for $\sigma\rightarrow\infty$.

The simulation results for the order parameter are shown as symbols in figure \ref{fig:phiMult_bifur}. At this scale no finite-size-effects can be seen at the phase transition. In a thorough data analysis with finite-size-scaling relations at $\sigma_c$, in the way we did in the first example, we found exponents of $b\approx c\approx 0.5$ to fit the order parameter and fluctuations (see fig.~\ref{fig:phiMult_simu}). These scaling relations imply, again in the thermodynamic limit, that the order parameter vanish and the fluctuations diverge as $m(\sigma)\sim(\sigma_c-\sigma)^{1/2}$ and  $\chi(\sigma)\sim|\sigma_c-\sigma|^{-\gamma}$, with $\gamma=c/b\approx 1$ respectively.

\begin{figure}[h]
 \centering
 \includegraphics[bb=50 50 276 276,width=6cm]{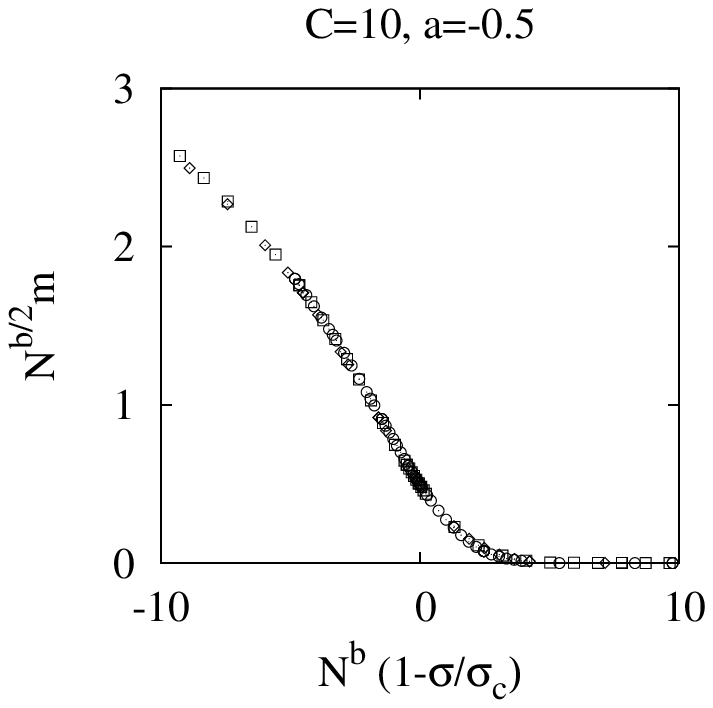}
 \includegraphics[bb=50 50 276 276,width=6cm]{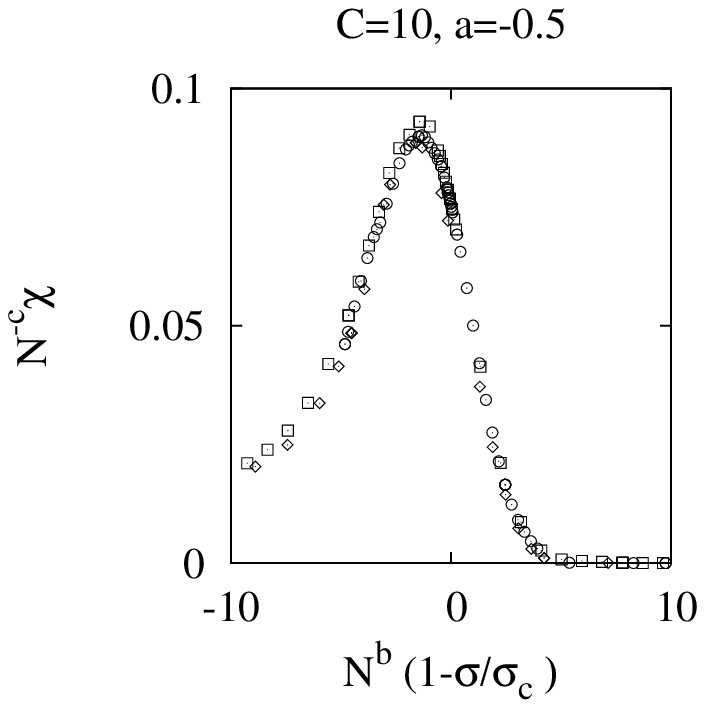}
\caption{Finite-size scaling analysis of the Landau-Ginzburg model with multiplicative noise  for $a=-0.5$, $C=10$. Rescaled ensemble average $m$ (\textit{left}) and fluctuations $\chi$ (\textit{right}) of $10^3$ numerical simulations with $N=10^3, 10^4, 10^5$ (circles, squares, diamonds). Critical point, as from Eq.~(\ref{eq:selfCons}), is $\sigma_c=2.169$; exponents: $b=0.5, c=0.5$.}
\label{fig:phiMult_simu}
\end{figure}

\subsection{Canonical model for noise-induced phase transitions}
\label{cha:canonicalModelPhaseTransition}

At the last example we will study a model for which a genuine phase transition induced by multiplicative noise has been shown \cite{BPT:1994,BPTK:1997} with the feature that the ordered phase is reentrant, it only exists for intermediate noise intensities. The equation for an individual element is:
\begin{equation}
 \label{eq:can_ODE}
	\dot x_i=-x_i\,\left(1+x_i^2\right)^2+\left(1+x_i^2\right)\eta_i+C\left(X-x_i\right) 
\end{equation}
and the reduced system according to section~\ref{cha:method} (again setting $\langle\eta_i\rangle=0$) reads: 
\begin{subeqnarray} 
\label{eq:OPE_canipt_system}
 \slabel{eq:OPE_X_noiseInducedPT}
	\dot{X}&=&-X  \left( 1+{X }^{2} \right) ^{2}+\frac{1}{2}\, \left[ -12\, \left( 1+{X }^{2} \right) X -8\,{X }^{3} \right] \Omega+2\,X W\\
 \slabel{eq:OPE_O_noiseInducedPT}
	\dot \Omega &=& 2\,\Omega\, \left[ -\left( 1+{X }^{2} \right) ^{2}-4\,{X}^{2} \left( 1+{X }^{2} \right) +C \right] +2\,W \left( 1+{X }^{2} \right) \\
 \slabel{eq:OPE_W_noiseInducedPT}
	\dot W&=& W \left[ - \left( 1+{X }^{2} \right) ^{2}-4\,{X }^{2} \left( 1+{X}^{2} \right) +C \right] +{\sigma} ^{2} \left( 1+{X }^{2} \right)\, .
\end{subeqnarray}
The equilibrium condition~(\ref{eq:equilibrium_condition})
 becomes
\begin{eqnarray}
\label{eq:can_equilibrium}
0=-X \left( 1+{X}^{2} \right) ^{2}+{\frac { \left( - \left(6+6\,{X}^{2} \right) X-4\,{X}^{3} \right) {\sigma}^{2} \left( 1+{X}^{2} \right) ^{2}}{ \left( 1+6\,{X}^{2}+5\,{X}^{4}+C \right) ^{2}}}+\\\nonumber
+2\,{\frac {X{\sigma}^{2} \left( 1+{X}^{2} \right)}{1+6\,{X}^{2}+5\,{X}^{4}+C}} \,.
\end{eqnarray}
Equation (\ref{eq:can_equilibrium}) has the stable solution $X=0$ for $\sigma<\sigma_c$ or a pair of symmetric solutions $X\neq0$ for $\sigma>\sigma_c$. Here $X=0$ becomes unstable~(see fig.~\ref{fig:can_bifur}). The value of $\sigma_c$ indicates the location of a second-order phase transition. It follows from analyzing the Jacobian of (\ref{eq:OPE_X_noiseInducedPT}-\ref{eq:OPE_W_noiseInducedPT}) and calculates to:
\begin{equation}
 \label{eq:can_sigmaC}
\sigma_{c}=\frac{1+C}{\sqrt{2C-4}} \, .
\end{equation}
Accordingly, a minimal coupling $C>2$ is necessary to induce the phase transition. An analysis of this relation shows that $\sigma_c$ has a minimum with respect to $C$. Therefore,  see figure  \ref{fig:can_sigmaDeC}, the transition is predicted to be reentrant with respect to $C$: the ordered phase only exists in a range of values for $C$, with the surprising prediction that too a large coupling destroys the ordered state. The predictions of the order parameter expansion are in qualitative agreement with those obtained after solving the self-consistency equation. However, whereas the order parameter expansion predicts incorrectly that the order parameter monotonously increases with $\sigma$, as shown in figure \ref{fig:can_bifur},
the self-consistency equation instead predicts that the transition is reentrant also with respect to the quenched noise intensity $\sigma$, see figure  \ref{fig:can_sigmaDeC}. Both reentrant behaviors were observed in the case of  time-dependent noise \cite{BPT:1994,BPTK:1997}.
\begin{figure}[h]
 \centering
 \includegraphics[width=5cm,bb=20 118 575 673]{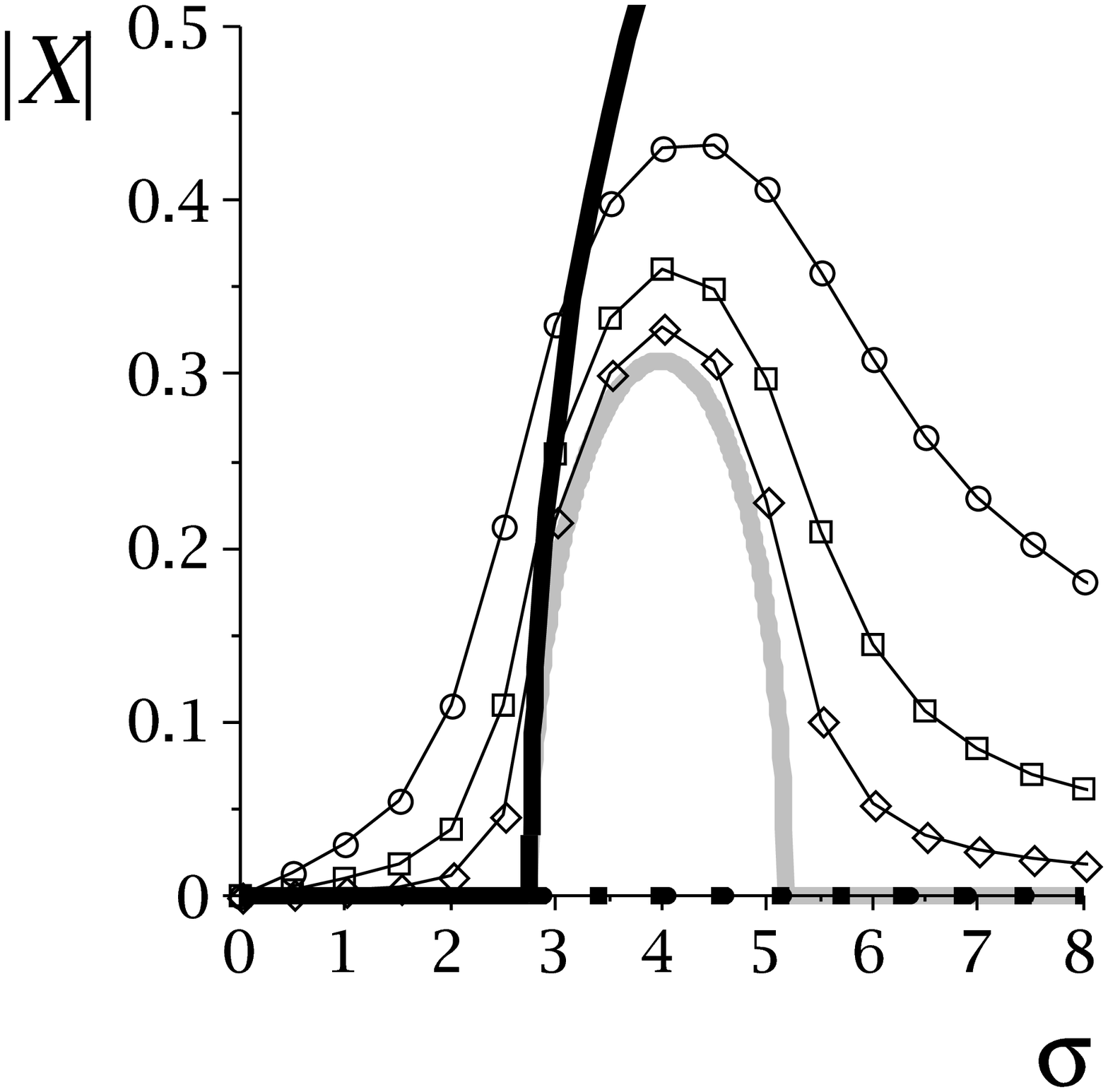}
\includegraphics[width=5cm,bb=20 118 575 673]{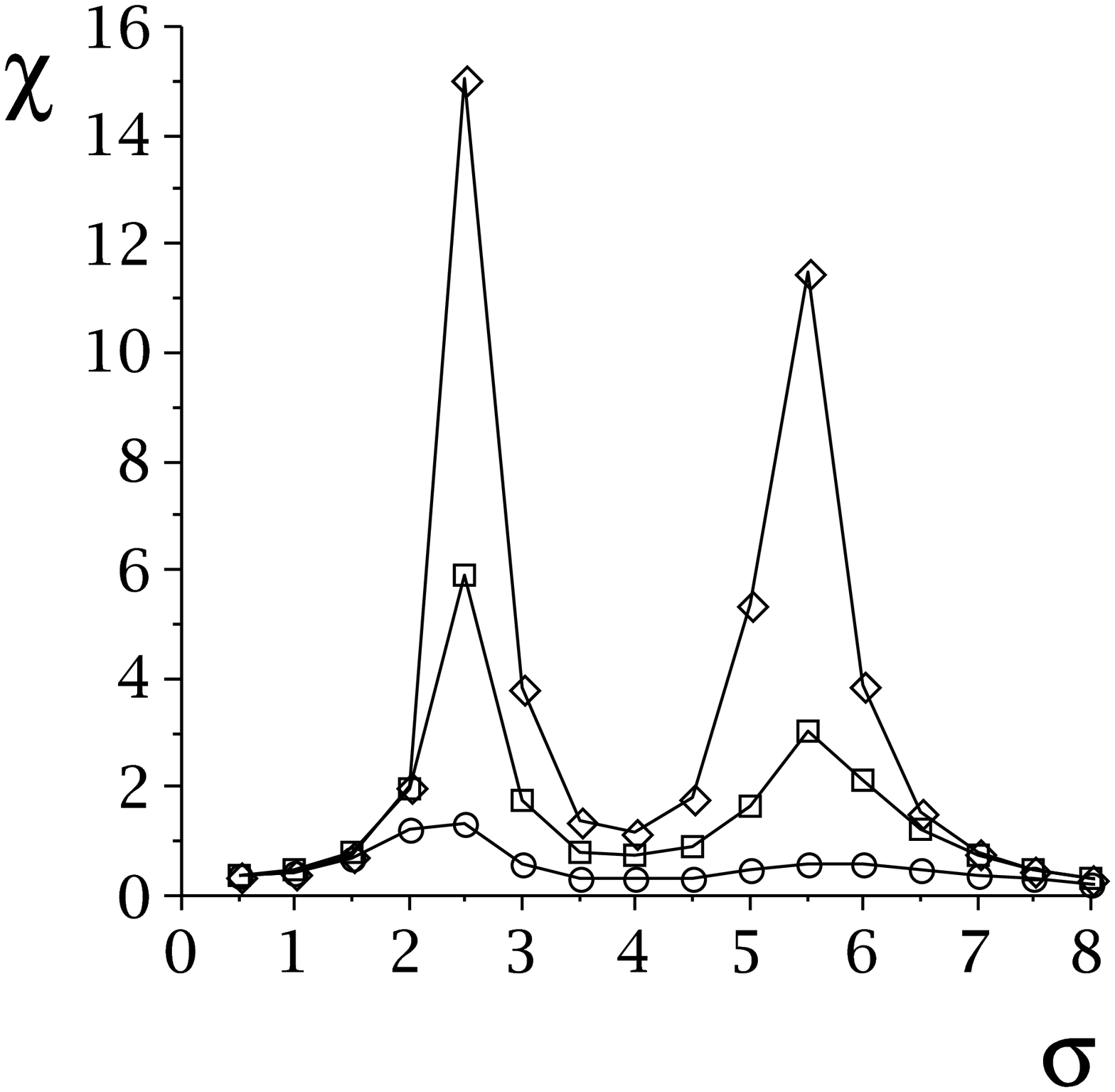}
\caption{Bifurcation diagram of model (\ref{eq:can_ODE}) (\textit{left}). Order parameter expansion (thick black line) and exact solution (grey line) together with the ensemble average  of $10^3$ numerical simulations for $N=10^3,10^4,10^5$ (circles, squares, diamonds). Coupling is $C=10$. On the \textit{right} the unscaled fluctuations are shown.
\label{fig:can_bifur}}
\end{figure}
\begin{figure}[h]
\centering
\includegraphics[width=5cm]{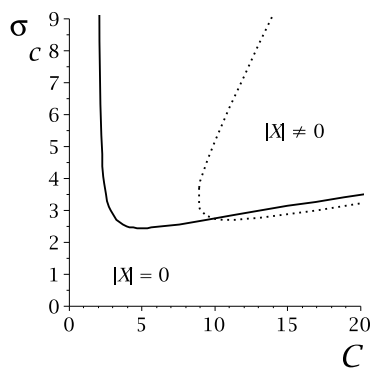}
\caption{Critical noise intensity versus coupling strength for model (\ref{eq:can_ODE}). The prediction (\ref{eq:can_sigmaC}) of the order parameter expansion (continuous line) and the exact solution (\ref{eq:selfCons}) as dotted line. Only the latter shows reentrance with respect to $\sigma$.
 \label{fig:can_sigmaDeC}}
\end{figure}

Again we have compared the predictions with the numerical integration of the set of equations~(\ref{eq:can_ODE}). The simulation results are shown as symbols in figure~\ref{fig:can_bifur}. Due to finite-size-effects the theoretical results are approached with increasing number $N$ of particles, reentrance and the dependence of $\sigma_c$ from $C$ are observed. Analyzing the data as we have done with the other examples, we find exponents for the scaling relations of $b\approx 0.33$ and $c\approx0.67$. As in the first case this implies the relations  $m(\sigma)\sim(\sigma_c-\sigma)^{1/2}$ and $\chi(\sigma)\sim|\sigma_c-\sigma|^{-\gamma}$, $\gamma=c/b\approx 2$ in the thermodynamic limit. Figure~\ref{fig:can_simu} summarizes the fitted simulation data.

\begin{figure}[h]
 \centering
 \includegraphics[bb=50 50 276 276,width=6cm]{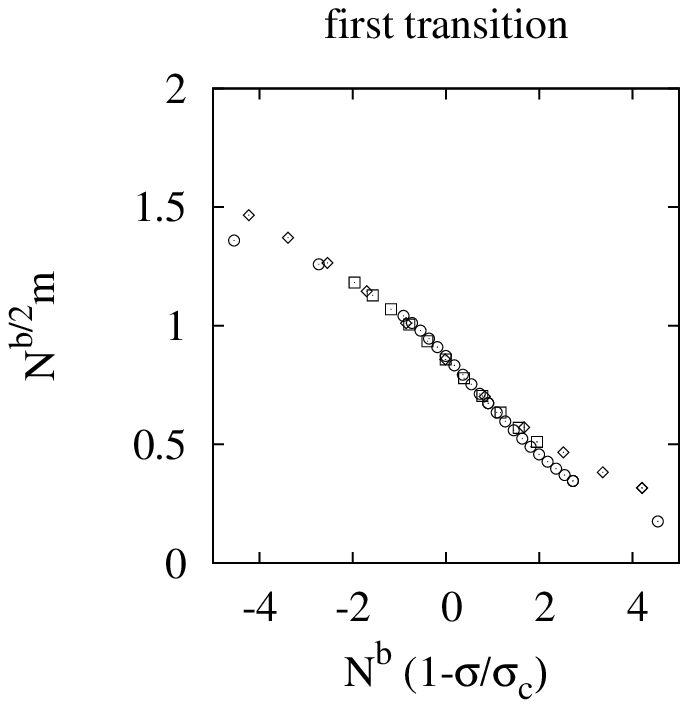}
 \includegraphics[bb=50 50 276 276,width=6cm]{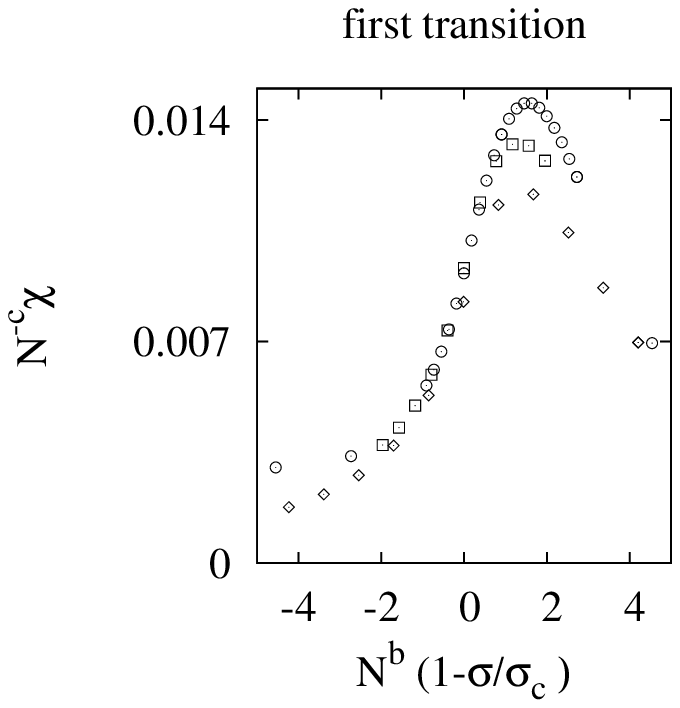}
 \includegraphics[bb=50 50 276 276,width=6cm]{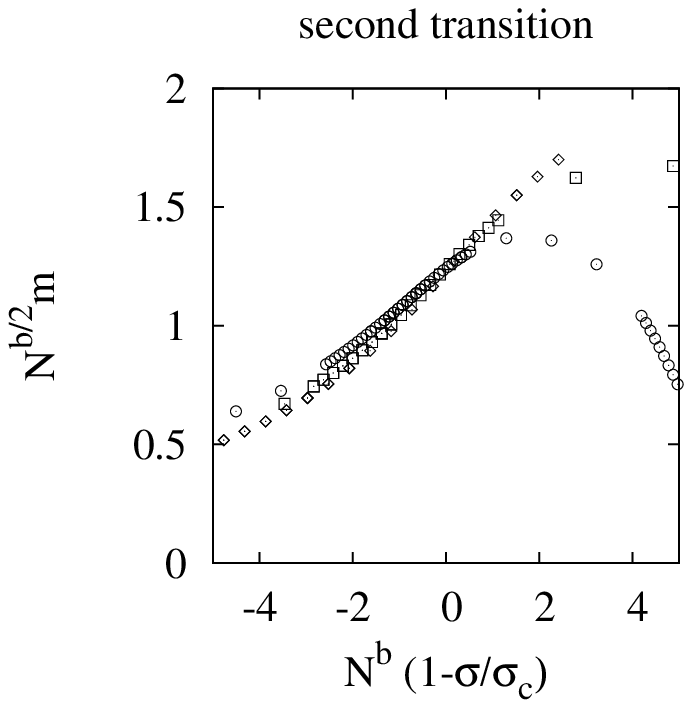}
 \includegraphics[bb=50 50 276 276,width=6cm]{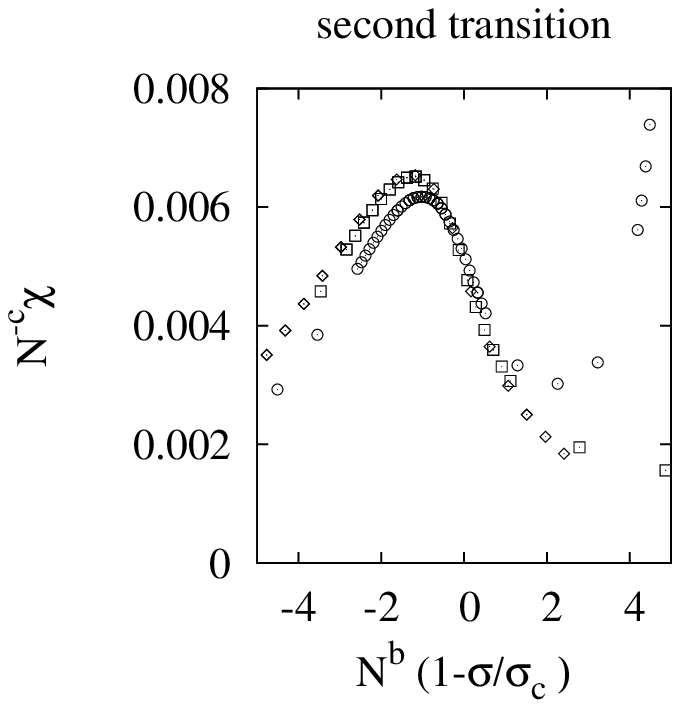}
\caption{Rescaled simulation data for the first (\textit{top}) and the second (\textit{bottom}) phase transition ($C=10$). Mean value (\textit{left}) and fluctuations (\textit{right}) for ensembles of $N=10^3,10^4,10^5$ (circles, squares, diamonds). Critical points, as from Eq.~(\ref{eq:selfCons}), are $\sigma_c=2.749$ for the first and $\sigma_c=5.169$ for the second transition. Exponents are: $b=0.33, c=0.67$.}
\label{fig:can_simu}
\end{figure}

\section{Conclusions}
\label{cha:discussion}

In this paper we have constructed an approximate analytical scheme based on the order parameter expansion~\cite{deMonteOvidio2002,deMonteOvidioMosekilde2003,deMonteOvidioChateMosekilde2004,deMonteOvidioChateMosekilde2005,iacyel2006} to study the macroscopic behavior of extended systems which are globally coupled. We have used the method to study in detail the phase diagram of three widely used models of phase transitions in scalar systems: the Landau-Ginzburg scalar model with both additive and multiplicative quenched noise and a genuine model for noise-induced phase transitions where time-dependent noise has been replaced by quenched, time-independent noise coupled multiplicatively to the dynamical variable~\cite{BPT:1994}.

We have compared the results of our simple approach with those coming from a numerically involved, but in principle exact, treatment based on the self-consistency relation and with extensive numerical simulations of the corresponding dynamical equations for each model. In the case of additive noise, the main result is that macroscopic order is destroyed when increasing the intensity of the quenched noise. In the other two cases, when noise appears multiplicatively, we find that macroscopic order appears for intermediate value of the intensity of the quenched noise. Since the quenched noise can represent, for instance, diversity or heterogeneity, it appears paradoxically that some amount of structural disorder is needed in order to observe macroscopic order.

Furthermore it has been shown numerically, that all investigated models follow a finite-size scaling law and the exponents have been determined. It suggests a common universality class for the Landau-Ginzburg model with additive quenched noise and the canonical model for noise induced phase transitions, whereas the Landau-Ginzburg model with multiplicative quenched noise yields different exponents. A more detailed analysis of the finite-size relations and their possible dependence with the system parameters will be presented elsewhere~\cite{NikoRaul:inProgress}.

The method of order parameter expansion, which we lead consistently up to terms of second order, is a tool which reduces large systems to  only a couple of reduced variables. The advantage is its very easy management. In this paper we have proven that reliable conclusions can be drawn with that method in some cases. Since it is an expansion around mean values the method yields good results for low values of the intensity of the quenched noise or for high synchronization of the subunits. Otherwise, the method might not be reliable. As a consequence the reentrant phase transitions were not predicted in the studied cases for multiplicative noise. It is an open issue how to modify the method in order to predict the reentrant transitions.

\section{Acknowledgements}
The authors acknowledge financial support from the EU NoE BioSim, LSHB-CT-2004-005137, and project FIS2007-60327 from MEC (Spain) and FEDER (EU). NK is supported by a grant from the Govern Balear.


\begin{thebibliography}{10}

\bibitem{GS:1999}
{J. Garc\'{\i}a-Ojalvo} and {J.M. Sancho}.
\newblock {\em Noise in Spatially Extended Systems}.
\newblock Springer--Verlag, New York, 1999.

\bibitem{srrmp}
L.~Gammaitoni, P.~H\"{a}nggi, P.~Jung, and F.~Marchesoni.
\newblock Stochastic resonance.
\newblock {\em Rev. Mod. Phys.}, 70:223, 1998.

\bibitem{HM:2009}
{P. H\"{a}nggi} and {F. Marchesoni, eds}.
\newblock Topical issue on stochastic resonance.
\newblock In {\em The European Physical Journal}, volume~69, 2009.

\bibitem{PK:1997}
{A. Pikovsky} and {J. Kurths}.
\newblock Coherence resonance in a noise-driven excitable system.
\newblock {\em Phys. Rev. Lett.}, 78:775, 1997.

\bibitem{ZGBUK:2003}
{A.A. Zaikin}, {J. Garc\'{\i}a-Ojalvo}, {R. B\'ascones}, {E. Ullner}, and {J.
  Kurths}.
\newblock Doubly stochastic coherence via noise-induced symmetry in bistable
  neural models.
\newblock {\em Phys. Rev. Lett.}, 90:030601, 2003.

\bibitem{SCSW:1997}
M. Santagiustina, P. Colet, M. San~Miguel, and D. Walgraef.
\newblock Noise-sustained convective structures in nonlinear optics.
\newblock {\em Phys. Rev. Lett.}, 79(19):3633--3636, Nov 1997.

\bibitem{CFT:2006}
M.G. Clerc, C.~Falc\'on, and E.~Tirapegui.
\newblock Front propagation sustained by additive noise.
\newblock {\em Phys. Rev. E}, 74(1):011303, Jul 2006.

\bibitem{BPT:1994}
{C. van den Broeck}, {J.M.R. Parrondo}, and {R. Toral}.
\newblock Noise-induced nonequilibrium phase transition.
\newblock {\em Phys.~Rev.~Lett.}, 73:3395, 1994.

\bibitem{BPTK:1997}
{C. van den Broeck}, {J.M.R. Parrondo}, {R. Toral}, and {R. Kawai}.
\newblock Nonequilibrium phase transitions induced by multiplicative noise.
\newblock {\em Phys. Rev. E}, 55:4084, 1997.

\bibitem{tessone2006}
C.J. Tessone, C.R. Mirasso, R.~Toral, and J.D. Gunton.
\newblock Diversity-induced resonance.
\newblock {\em Phys. Rev. Lett.}, 97:194101, 2006.

\bibitem{tessone2007a}
C.J. Tessone, A.~Scir{\`e}, R. Toral, and P.~Colet.
\newblock Theory of collective firing induced by noise or diversity in
  excitable media.
\newblock {\em Physical Review E}, 75, 2007.

\bibitem{ToralTessoneLopes2007}
R.~Toral, C.J. Tessone, and J.V. Lopes.
\newblock Collective effects induced by diversity in extended systems.
\newblock {\em Eur. Phys. J. Special Topics}, 143:59--67, 2007.

\bibitem{toral2008a}
R.~Toral, E.~Hern\'andez-Garc{\'\i}a, and J.D. Gunton.
\newblock Diversity-induced resonance in a system of globally coupled linear
  oscillators.
\newblock {\em International Journal of Bifurcations and Chaos}, 19:3499, 2009.

\bibitem{chen2008}
H.~Chen and J.~Zhang.
\newblock Diversity-induced coherence resonance in spatially extended chaotic
  systems.
\newblock {\em Physical Review E}, 77, 2008.

\bibitem{Gosak2009}
M.~Gosak.
\newblock Cellular diversity promotes intercellular ca2+ wave propagation.
\newblock {\em Biophysical Chemistry}, 139:53, 2009.

\bibitem{Ullner2009}
E.~Ullner, J.~Buceta, A.~Diez-Noguera, and J.~Garcia-Ojalvo.
\newblock Noise-induced coherence in multicellular circadian clocks.
\newblock {\em Biophysical Journal}, 96:3573, 2009.

\bibitem{Zanette:2009}
D.~Zanette.
\newblock Interplay of noise and coupling in heterogeneous ensembles of phase
  oscillators.
\newblock {\em European Physical Journal B}, 69:269, 2009.

\bibitem{Tessone2009}
C.J. Tessone and R.~Toral.
\newblock Diversity-induced resonance in a model for opinion formation.
\newblock {\em European Physical Journal B}, 71:549, 2009.

\bibitem{Postnova09}
S.~Postnova, K.~Voigt, and H.A. Braun.
\newblock A mathematical model of homeostatic regulation of sleep-wake cycles
  by hypocretin/orexin.
\newblock {\em Journal of Biological Rhythms}, 24:523, 2009.

\bibitem{Chen09}
H.S. Chen, Y.~Shen, and Z.H. Hou.
\newblock Resonant response of forced complex networks: The role of topological
  disorder.
\newblock {\em Chaos}, 19:033122, 2009.

\bibitem{Wu09}
D.~Wu, S.Q. Zhu, and X.Q. Luo.
\newblock Cooperative effects of random time delays and small-world topologies
  on diversity-induced resonance.
\newblock {\em European Physics Letters}, 86:50002, 2009.

\bibitem{Perc08}
M.~Perc, M.~Gosak, and S.~Kralj.
\newblock Stochastic resonance in soft matter systems: combined effects of
  static and dynamic disorder.
\newblock {\em Soft Matter}, 4:1861, 2008.

\bibitem{Acebron07}
J.A. Acebron, S.~Lozano, and A.~Arenas.
\newblock Amplified signal response in scale-free networks by collaborative
  signaling.
\newblock {\em Physical Review Letters}, 99:128701, 2007.

\bibitem{TSTP:2007}
C.J. Tessone, A.~Scire, R.~Toral, and P.~Colet.
\newblock Diversity-induced resonance.
\newblock {\em Phys. Rev. E}, 75:016203, 2007.

\bibitem{deMonteOvidio2002}
S.~{de Monte} and F.~d'Ovidio.
\newblock Dynamics of order parameters for globally coupled oscillators.
\newblock {\em Europhys. Lett.}, 58:21--27, 2002.

\bibitem{deMonteOvidioMosekilde2003}
S.~{de Monte}, F.~d'Ovidio, and E.~Mosekilde.
\newblock Coherent regimes of globally coupled dynamical systems.
\newblock {\em Phys. Rev. Lett}, 90(5):054102, 2003.

\bibitem{deMonteOvidioChateMosekilde2004}
S.~{de Monte}, F.~d'Ovidio, Hugues Chat{\'e}, and E.~Mosekilde.
\newblock Noise-induced macroscopic bifurcations in globally coupled chaotic
  units.
\newblock {\em Phys. Rev. Lett.}, 92:254101, 2004.

\bibitem{deMonteOvidioChateMosekilde2005}
S.~{de Monte}, F.~d'Ovidio, Hugues Chat{\'e}, and E.~Mosekilde.
\newblock Effects of microscopic disorder on the collective dynamics of
  globally coupled maps.
\newblock {\em Physica D}, 205:25--40, 2005.

\bibitem{iacyel2006}
I.~Gomes~Da Silva, S.~{de Monte}, F.~d'Ovidio, R. Toral, and C.~R.
  Mirasso.
\newblock Coherent regimes of mutually coupled chua‚s circuits.
\newblock {\em Physical Review E}, 73(036203), 2006.

\bibitem{msmtoral00}
M.~San Miguel and R.~Toral.
\newblock Stochastic effects in physical systems.
\newblock In J.~Martinez, E.~Tirapegui and R.~Tiemann, editors, {\em
  Instabilities and nonequilibrium structures VI}, pages 35--120. Kluwer
  academic publishers, 2000.

\bibitem{stanley}
H.E. Stanley.
\newblock {\em Introduction to phase transitions and critical phenomena}.
\newblock Oxford university press, 1971.

\bibitem{landaubinder}
D.P. Landau and K.~Binder.
\newblock {\em A guide to Monte Carlo simulations in statistical physics}.
\newblock Cambridge university press, 2000.

\bibitem{amit}
D.J. Amit and V.M. Mayor.
\newblock {\em Field Theory, the Renormalization Group and Critical Phenomena}.
\newblock World Scientific Publishing Co.Pte. Ltd., 3rd edition, 2005.

\bibitem{young}
A.P. Young, editor.
\newblock {\em Spin Glasses and Random Fields}.
\newblock World Scientific Publishing Co.Pte. Ltd., 1998.

\bibitem{cardy}
J.L. Cardy.
\newblock {\em Finite-Size Scaling}.
\newblock Elsevier science publishers, 1988.

\bibitem{deutsch:92}
H.P. Deutsch.
\newblock Optimized analysis of the critical behavior in polymer mixtures from
  monte carlo simulations.
\newblock {\em J. Stat. Phys}, 67:1039, 1992.

\bibitem{VPAH94}
C.~{Van den Broeck}, J.M.R. Parrondo, J.~Armero, and A.~Hern{\'a}ndez-Machado.
\newblock Mean field model for spatially extended systems in the presence of
  multiplicative noise.
\newblock {\em Phys. Rev. E}, 49(4):2639--2643, Apr 1994.

\bibitem{GPSV96}
J.~Garc{\'\i}a-Ojalvo, J.M.R. Parrondo, J.M. Sancho, and C.~{Van den Broeck}.
\newblock Reentrant transition induced by multiplicative noise in the
  time-dependent ginzburg-landau model.
\newblock {\em Phys. Rev. E}, 54(6):6918--6921, Dec 1996.

\bibitem{BP:01}
J.M.R.~Parrondo, J.~Buceta and F.J. de~la Rubia.
\newblock Random ginzburg-landau model revisited: Reentrant phase transition.
\newblock {\em Physical Review E}, 63:031103, 2001.

\bibitem{StrogatzNonlinearDyn}
S.H. Strogatz.
\newblock {\em Nonlinear dynamics and chaos: with applications to physics,
  biology, chemistry and engineering}.
\newblock Addison-Wesley, 1994.

\bibitem{NikoRaul:inProgress}
N.~Komin and R. Toral.
\newblock work in progress.

\end{thebibliography}
\end{document}